\documentclass[prl,aps,showpacs,amsmath,amssymb,superscriptaddress,twocolumn]{revtex4}
\usepackage{graphicx}
\usepackage{dcolumn}
\usepackage{bm}
\usepackage{subfigure}
\usepackage{setspace}
\usepackage{feynmp}
\usepackage{psfrag}
\newcommand{\ket}[1]{\ensuremath{| #1 \rangle}}
\DeclareMathSizes{10}{9}{6}{6}
\begin{document}

\title{Spontaneously broken gauge symmetry in a Bose gas with constant particle number}
\author{Alexej Schelle}
\affiliation{A. Schelle, Marienbader Stra\ss e 2, \\ 93057 Regensburg, email: alexej.schelle@lycos.com}
\date{\today}

\begin{abstract}
The interplay between spontaneously broken gauge symmetries and Bose-Einstein condensation
has long been controversially discussed in science, since the equation of motions are invariant under phase transformations.
Within the present model it is illustrated that spontaneous symmetry breaking appears as a non-local process in position space, but within disjoint subspaces of the underlying Hilbert space. 
Numerical simulations show that it is the symmetry of the relative phase distribution between condensate and non-condensate quantum fields which is spontaneously broken when passing the critical temperature for Bose-Einstein condensation. 
Since the total number of gas particles remains constant over time, the global U(1)-gauge symmetry of the system is preserved. 
\end{abstract}
\maketitle

\section{1. Introduction}

It is hard to find perfect symmetries in daily life, if one puts focus on classical objects.
One may easily recognize the asymmetry of a seemingly "perfect" melon, i.e. 
$\sqrt{x^2+y^2+z^2}=R$, if one analyzes the $\mathbb{R}^3$ symmetry structure 
of the melon e.g. by comparing slices in the $\mathbb{R}^2$ plane with circles obeying 
$\sqrt{x^2+y^2}=r$, where 
$r>0$ is the (variable) radius of a slice.
In quantum mechanics, formally, it is rather straight forward to define a perfect symmetry \cite{Wigners}. 
Let $\ket{\Psi}$ be a wave function (wave field in bracket notation). 
Then the operation $\mathcal{T}$ defines a symmetry operation on the state $\ket{\Psi}$,
if $\mathcal{T}\ket{\Psi}$ is norm-preserving and leaves the laws of physics of a given system invariant under this symmetry operation $\mathcal{T}$. 
As an example, any (closed) quantum system obeying Schr\"{o}dinger's equation is rotationally symmetric under rotations $\mathcal{R}$,  
since the transformation $\Psi(r)\rightarrow\mathcal{R}\Psi(r)$ with $\mathcal{R}\Psi(r):={\rm e}^{i\phi}\Psi(r)$ is (a) norm-preserving and 
(b) leaves the physical equations of state invariant \cite{Diracs}.

Processes inducing transitions from states of gauge symmetry to asymmetric states 
are called symmetry breaking.
If one considers a classical-to-quantum transition 
in an unrealistic, but helpful gedanken experiment one comes to the conclusion that the 
transition from a symmetric state to a asymmetric state (symmetry breaking) 
is likely to depend on the scales of the system. 
Imagine two points placed at a relative distance $d$ on top of the melon. 
When the melon is rotated by an angle $\phi$ in the $\mathbb{R}^3$ plane without changing the $z$ positions of the points
(e.g. using a norm-preserving rotation $\mathcal{R}_\phi\in \mathcal{S}\mathcal{O}(3)$ in order to rotate the melon), the distance is 
not preserved, because of the underlying asymmetry of the melon. 
Thus, the physics of the system depends on the choice of the angle $\phi$ 
(e.g. if the points are charged), 
because of the underlying melon symmetry or asymmetry, respectively.
Taking a continuous limit to the quantum regime, i.e. $d, R\rightarrow0^+$ and 
temperature $T\rightarrow0^+$, the change in distance induced by asymmetry 
becomes on the order of the particles' uncertainties, i.e. their wave lengths, 
and thus the distance remains 
preserved within the resolution of the quantum limit.
Indeed, in this (quantum) limit the symmetry of the system gets independent 
on the choice of the angle (phase) $\phi$ (compare e.g. Schr\"{o}dinger's equation \cite{Tan_Dup_Gry}).

Bose-Einstein condensates are particularly interesting for analyses of broken gauge symmetries, since they reflect a natural classical-to-quantum transition from a microscopic quantum 
(statistical) state to a macroscopic state of matter which obeys both the 
characteristics of quantum mechanics induced by uncertainty (particle wave duality) and 
the laws of classical Boltzmann statistics for indistinguishable particles in the semi-classical limit. 
Switching back to well-founded mathematical theorems, 
it has been shown that spontaneous gauge symmetry breaking is a necessary and sufficient 
process for Bose-Einstein condensation.
A clear understanding of the interplay of spontaneous gauge symmetry breaking and Bose-Einstein condensation \cite{Ketterle, Legg_Sols, Wagner, Elizur, Broken} is still of current scientific interest.
In this context, it has e.g. long been unclearified whether the absolute phase of a Bose-Einstein condensate is completely random or whether it is not well-defined, since the total average 
of the quantum field should always be zero \cite{Laloe, Legg_Sols, Oberthaler, Cons1, Cons2} as long as the total number of particles is conserved.
Is it possible that the absolute phase of a Bose-Einstein condensate has a pre-defined alignement due to spontaneous symmetry breaking? 

\section{2. Theory}

In this Article, it is illustrated for a Bose gas with constant particle number that the absolute phase of the Bose gas is completely random. However, a broken gauge symmetry arises below the critical temperature as asymmetry of the relativ phase distribution between condensate and non-condensate field, thus in disjoint subspaces of the underlying Hilbert spaces.
The field modes are numerically analysed modelling the underlying complex valued fugacity spectrum, considering very weak (s-wave) interactions and non-classical correlations between the particles. 
The phase gauge symmetry breaking process is continuously monitored by 
switching the gas temperature from above the critical temperature close to zero, and drawing (average) realizations of the condensate and non-condensate quantum field, which 
is treated as a coherent field with randomly (Boltzmann distributed) field mode 
occupation numbers using a random (Markov chain) Monte Carlo Metropolis algorithm.   
The spectrum implies a macroscopically and locally broken phase gauge symmetry of the average condensate and non-condensate quantum field, and the results indicate that the broken gauge symmetry can in principle be experimentally verified by measuring the local phase distribution 
of the condensate part of the Bose gas.
Since the total i.e. global U(1)-gauge symmetry of the physical system is consistently preserved when passing the ideal gas critical tempature for Bose-Einstein condensation (particle number conservation), 
the global phase of the Bose-Einstein condensate is well-defined, but totally random.

From the symmetry definition of the Introduction, it is straight forward to 
understand that Noether's theorem \cite{Noether} 
can be understood as the classical pendant, or even a precursor 
of the symmetry definition of quantum mechanics.
It states that 
any continuous symmetry implies a conserved quantity. 
Noether's theorem furthermore implies that, if a physical quantity is not conserved over time,
then there can be no underlying continuous symmetry.
Thus, since the global phase and the total number of particles of a Bose-Einstein condensate are conjugate variables, this implies that the process of gauge symmetry breaking may be due to gain or loss of gas particles. 
If one keeps the number of gas particles constant over time, particles can only be exchanged between the two formal subsystems condensate and non-condensate, so that only
the phase gauge symmetry of the relative phase distribution between these two subsystems may be broken below the critical temperature, since the global phase gauge symmetry of the condensate is preserved and thus the phase is absolutely random.
Thus, spontaneous symmetry breaking may only effect on the relative phase distribution between condensate and non-condensate subsystems.

In order to put more focus on this conjecture, let's recover some
number-conserving theories
of Bose-Einstein condensation \cite{NCMET, Cons1, Cons2}.
Since all equations of state can be defined in terms of 
quantum fields, we consider the decomposition

\begin{equation}
\hat{\Psi}(\textbf{r}) = \hat{\Psi}_0(\textbf{r}) + \hat{\Psi}_\perp(\textbf{r})
\end{equation}\\
in second quantization, and furthermore the spatial and number average of the 
wave field 

\begin{equation}
\psi = \int \bf{d} \bf{r}\langle\hat{\Psi}(\textbf{r})\rangle = \int \bf{d} \bf{r} \langle\hat{\Psi}_0(\textbf{r}) \rangle + \int d \bf{r} \langle\hat{\Psi}_\perp(\textbf{r}) \rangle .
\end{equation}\\
It is Elizur's theorem \cite{Elizur} which motivates the definition of the order paramter in the above way, since the theorem implies that order parameters must be non-local objects.
Indeed, position dependend single particle wave functions can not change their symmetry properties during Bose-Einstein condensation spontaneously for sufficiently weak interactions, thus in order to find spontaneously broken gauge symmetries it is only important to analyse the global spatial average of the quantum fields. 
Processes which may lead to a spontaneuous broken gauge 
symmetry inducing (many body) phase coherence \cite{Glauber} are non-classical correlations (entanglement) between condensate and non-condensate particles, which implies formally local particle-number breaking (compare e.g. \cite{Cons2, thesis, Gardiner}). 

The Heisenberg equation of state is invariant under phase transformations 
$\psi\rightarrow\psi{\rm e}^{i\phi}$ and thus implies a 
continuous phase gauge symmetry \cite{Heisenberg}. 
Since the number of particles in the Bose gas is conserved,

\begin{equation}
\psi  = 0
\end{equation}\\
at all times. Any non-zero field average would contradict particle number conservation and 
correspond to a globally and spontaneously broken gauge symmetry. 
Thus, the global U(1)-gauge symmetry remains preserved. 
One could argue that the symmetry breaking occurs on time scales smaller than the energy scales defined by the energy uncertainty.
However, those processes only occur as vanishingly small imaginary parts on 
the diagonal part of the density matrix 
and thus do not physically contribute on the energy scale of the considered system \cite{thesis}. 
On the other hand, if one argues that also the total number of particles may fluctuate, 
symmetry breaking due to these fluctuations cannot occur in subsystems of conserved particle number, but only in non-classically correlated total number states.
Since number fluctuations of the total particle number are naturally 
classical without coherences between different (total particle) number states, 
spontaneous symmetry breaking may thus not occur in a Bose-Einstein condensate with uncertain total number of particles - which is mathematically well-defined by an ensemble of subsystems with constant particle number.
Hence, 

\begin{equation}
\int \bf{d} \bf{r}\langle\hat{\Psi}_0(\textbf{r})\rangle = - \int \bf{d} \bf{r}\langle\hat{\Psi}_\perp(\textbf{r})\rangle \ ,
\label{shift}
\end{equation}\\
which means that the invariance of $\psi = 0$ under norm preserving rotations implies that the 
particle number is conserved (Noether's theorem), and vice versa, a constant particle number indicates that the symmetry of the total 
field is plausible and hence gauge symmetries can only be broken in different subsystems of the underlying Hilbert space.

\begin{figure}[t]
\includegraphics[width=9.0cm, height=7.0cm,angle=0.0]{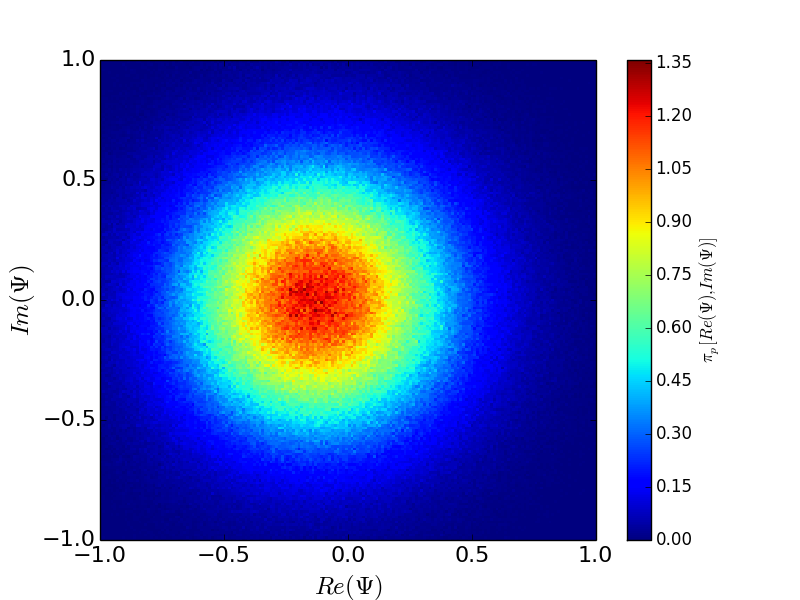} 
\caption{(color online) Real and Imaginary parts of $4\times10^6$ realizations 
of the average non-condensate random field in Eq. (\ref{NC_field}) for $N=1000$ particles in a trap with trap frequencies $\omega_x,\omega_y = 2\pi\times42.0$ Hz, 
$\omega_z = 2\pi\times120.0$ Hz below ($T = 1.0$ nK) the ideal gas critical 
temperature $T_c = 26.9$ nK. 
Number of considered eigen modes in the trap is $n_{\rm cut}=500$ corresponding to a 
trap depth of $2.8$ $\mu$K.
The probability distribution $\pi_p[{\rm Im}(\psi_{\perp}), {\rm Re}(\psi_{\perp})]$  
highlights the broken symmetry of the quantum field -- the (left) 
shifted real parts of the distribution.}
\label{figone}
\end{figure}

\section{3. Results}
In the subsequent section, numerical results for symmetry breaking processes 
are shown for Bose-Einstein condensates in three dimensions. 
The randomness of the condensate and non-condensate quantum field passing the critical temperature 
for Bose-Einstein condensation is modelled by different numerical realisations of the wave functions with random samples using a standard Monte Carlo Metropolis algorithm. 
Random fluctuations of the quantum field induced by temperature 
are accounted for using Boltzmann probability factors \cite{Krauth}.
For this purpose, the concept of imaginary time \cite{Abrikosov} is used. 
In this case, one realization of the random non-condensate field average 
can be mathematically expressed as

\begin{equation}
\psi_0 = -\psi_\perp = \int \bf{d} \bf{r}\langle\hat{\Psi}_\perp(\textbf{r})\rangle = \mathcal{N}^{-1}\sum_{k\ne0}c_k z(\mu^c_k) \ , 
\label{NC_field}
\end{equation}\\
where $c_k$ are (real valued) random probability amplitudes 
for a particle to occupy a state corresponding to the fugacity 
$z(\mu^c_k)=e^{\beta\mu^c_k}$, and $\mu^c_k$ is a complex valued chemical potential (of the $k^{th}$ state of the non-condensate), 
i.e. defined by the $k^{th}$ complex root of the (approximate) equation

\begin{equation}
N-N_\perp = \sum_{j\ne0}z^j(\mu^c)\left[\prod_{k=x,y,z}\frac{1}{1-e^{-j\beta\hbar\omega_k}}-1\right] 
\label{ptn}
\end{equation}\\
for a Bose-Einstein condensate in a harmonic trap with trapping frequencies 
($\omega_x, \omega_y, \omega_z$) and 
very weak s-wave (short range) interactions between the particles \cite{Legg_Sols, thesis, comment2}.
This implies a relative phase of $\pi$ between the subsystems condensate and non-condensate, while the  field equations remain gauge symmetric.
It can be shown that probability amplitudes moreover satisfy $c_k*c_l=\delta_{kl}$, because of the orthogonality relation which holds for single particle wave functions with sufficiently weakly interacting particles.
The probability for the field to occupy a state with random (number) amplitudes $\lbrace c_k\rbrace = \lbrace0..1\rbrace$ is given 
by the Boltzmann factor defined by the absolute value of the chemical potential $\mu$ 
and the complex valued inverse temperature,
 
 \begin{figure}[t]
\includegraphics[width = 7.0 cm, height = 5.0 cm,angle=0.0]{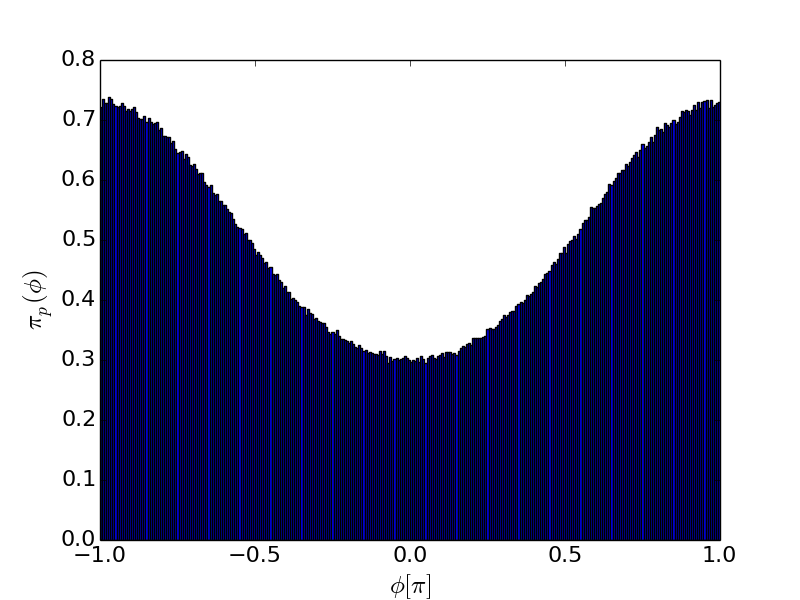} \newline                  
\caption{(color online) Phase distribution of the non-condensate quantum field obtained from 
the field distribution of the Bose-Einstein condensate in Fig. \ref{figone}.
Condensate phase distribution corresponds to shifting $\phi \rightarrow \phi \pm \pi = \phi_0$.
The non-condensate (and condensate) phase is distributed around 
$\pm\pi (0)$.}
\label{fig_fugacity}
\end{figure}

\begin{equation}
p(\mu)=\frac{e^{-\beta \lvert\mu\rvert}}{\mathcal{N}_\mu} ,
\label{prob}
\end{equation}\\
with $\mathcal{N}_\mu=\int_{\mathbb{C}}{\rm d}\mu{\rm~e}^{-\beta\lvert\mu\rvert}$ and 
$\beta = -it/\hbar$.
The total chemical potential of the state is given by 

\begin{equation}
\mu = {\rm Re}\lbrace\mu\rbrace + i\cdot{\rm Im}\lbrace\mu\rbrace = \sum_{k\ne0}\lvert c_k\rvert^2\mu_k^c
\end{equation}\\
Finally, it is easy to show that $\mathcal{N} = (\sum_{k\ne0}\lvert z(\mu^c_k)\rvert^2\lvert c_k\rvert^2)^{1/2}$.

The spectrum of the fugacity consists of two parts, a rotationally symmetric ring in the complex plane, 
which defines a quasi-continuum of metastable states in particular 
including the Boltzmann equilibrium for ${\rm Re}(z) \rightarrow 1$ and ${\rm Im}(z) \rightarrow 0$,
and a real valued symmetry breaking part which ranges 
from ${\rm Re}(z) \rightarrow 0^{+}$ and ${\rm Im}(z) = 0$ to the Boltzmann equlibrium 
${\rm Re}(z)=1, {\rm Im}(z) = 0$. 
Hence, 

\begin{equation}
\lvert\psi_{\perp}\rvert^2 = 1\ ,
\end{equation}\\
for any realization of the non-condensate field for which the probability amplitudes obey perfect reflection asymmetry, i.e. $c_k=-c_l$ with $\phi_{k}=\phi_l\pm\pi$, and 
 
 \begin{equation}
 \lvert\psi_{\perp}\rvert^2 = 0\ ,
 \end{equation}\\
if there is perfect reflection symmetry, i.e. $c_k=c_l$ for 
$\phi_{k}=\phi_l\pm\pi$ above the critical temperature. 

\section{4. Discussion}
This structure of the spectrum explains that the 
symmetric contributions of the random (non-condensate) quantum field 
(which build a circle in the complex plane) cancel out on average below the critical temperature, 
while the "symmetry breaking part" (${\rm Im}(z) = 0$) of the spectrum leads to a non-zero averages of the non-condensate quantum field.
Note that the non-condensate quantum field can be seen as imprint of the condensate field, since
the sum of the condensate and non-condensate quantum field preserves the global U(1)-gauge symmetry.
For temperatures approaching the critical temperature, the gauge symmetry of the non-condensate (and condensate) field is relatively (in the sense of two disjoint subspaces of the underlying Hilbert spaces) broken, so that

\begin{equation}
\overline{\psi_0} =-\overline{\psi_\perp} \ne 0 \ , 
\end{equation}\\
where the overline denotes different random (numerical) realizations of the quantum fields.
Our findings show that the spectrum of the fugacity for the non-condensate part of a Bose-Einstein condensate in a trap is
gapless as the real valued spectrum (projection onto the real parts) is. 
The imaginary parts of the fugacity entail the widths (inverse life times) of non-condensate many body states. 

In Fig. \ref{figone}, a number of $4\times10^6$ realizations of the non-condensate field is shown in order to illustrate symmetry breaking in the non-condensate (and condensate) field, which is figured using samples over Eqs. (\ref{NC_field}, \ref{ptn}) using the probability distribution 
in Eq. (\ref{prob}). 
The result shows a disc in the $({\rm Re}(\psi_\perp), {\rm Im}(\psi_\perp))$ complex 
plane which is (left) shifted in $x$ (real part) direction, 
because of the underlying microscopic asymmetry of the phase distribution for the non-condensate field (entailed in the non-condensate chemical potential).
As the symmetry of the fugacity spectrum is broken along 
the real axis in the complex plane, the average of the imaginary 
parts remains zero, whereas the average over 
the wave fields' real parts become non-zero.
These shifts lead to non-zero real valued and thus stable configurations 
of the average field, as highlighted by the probability distribution 
$\pi_p[{\rm Re}(\psi_\perp), {\rm Im}(\psi_\perp)]$.

The relative phase distribution of the non-condensate and condensate field in Fig. \ref{figone}, 
respectively, is defined by the angles 

\begin{equation}
\phi=\phi_0 + \pi \ , 
\end{equation}\\
where $\phi_0$ is the phase of the condensate average field.
The non-condensate (and condensate) phase is distributed around $\pm\pi (0)$, respectively, with a background $\pi_p$ of around $\pi_p\sim0.3$.
Thus, this implies that also the choice of the absolute phase of the condensate is totally random, but 
the relativ phase gauge symmetry between condensate and non-condensate field is spontaneously 
broken below the critical temperature.

Repeating the same calculus for different temperatures shows that the symmetry breaking 
part of the quantum field continuously tends from zero for large temperatures 
to negativ values as the temperature approaches zero.
As a consequence of the $U(1)$-symmetry breaking part of the spectrum, 
the average non-condensate (and condensate)
random field becomes non-zero as the temperature approaches zero, while the global 
$U(1)$-gauge symmetry and thus the constraint of 
total particle number conservation is preserved for all temperatures.
For temperatures larger than the critical temperature, the fugacity spectrum is no longer simply connected, but gapless. 
The limit $T\rightarrow \infty$ leads to a symmetric spectrum and therefore to a vanishing
of the broken phase gauge symmetry. 

\section{5. Conclusion}

At first glance, it may seem counter intuitive that the phase symmetry is broken below the critical temperature when comparing our results e.g. to the "melon gedanken experiment", since the
condensation process occurs onto the ground state of the Bose gas, which is symmetric. 
Thus, one may expect that the symmetry is preserved or even revealed below the critical temperature.
However, as illustrated in the analysis of this article, this is only a part of the entire condensate process, since it is the relative phase distribution of the condensate and non-condensate phase which matters.

Mathematically speaking, spontaneous symmetry breaking during Bose-Einstein condensation appears as non-zero real valued averages of complex number values representing non-condensate and condensate quantum fields, which highlight a asymmetric relative phase distribution of the two fields.
Thus, the origin of this shift can be assigned to the underlying microscopic symmetry structure of a finite set of two dimensional complex valued roots defined by the constraint that the sum of the two fields remains zero.

From the physical point of view, the process of gauge symmetry breaking during Bose-Einstein condensation occurs as a dynamical process in time (and temperature, respectviely) inducing phase coherence between condensate and non-condensate particles on a macroscopic scale when slowly cooling the gas temperature below the critical temperature expected for Bose-Einstein condensation.
Symmetry breaking during Bose-Einstein condensation 
can thus be explained as coherent interactions due to (particle-number breaking) scattering processes between condensate and non-condensate particles, which imprint non-random and quantum mechanically correlated phase shifts between the condensate and non-condensate particles inducing many-body coherence and thus non-zero field averages below the critical temperature. 

The author acknowledges \textit{fruitful} discussions with Andreas Buchleitner, Dominique Delande, Christopher Gaul, Benoit Gr\'{e}maud, Cord M\"{u}ller and Thomas Wellens on Bose-Einstein condensates.

\end{document}